\documentclass[iop]{emulateapj-rtx4}
\usepackage{bm}
\usepackage{threeparttable}
\usepackage{CJK}

\def\bfnabla{{\mbox{\boldmath $\nabla$}}}

\def\msun{M_\odot}

\renewcommand\bv{{\mbox{\boldmath $v$}}}
\newcommand\bb{{\mbox{\boldmath $B$}}}
\newcommand\bP{{\mbox{\boldmath $P$}}}
\newcommand\bF{{\mbox{\boldmath $F$}}}
\newcommand\bfr{{\sf\boldmath f}}
\newcommand\bI{{{\sf\boldmath I}}}

\def\<{\,\langle\langle}
\def\>{\,\rangle\rangle}

\begin{document}
\begin{CJK*}{UTF8}{gbsn}

\shortauthors{Y.-F. Jiang et al.}
\author{Yan-Fei Jiang (姜燕飞)\altaffilmark{1}, James M. Stone\altaffilmark{1} \& Shane W. Davis\altaffilmark{2}}
\affil{$^1$Department of Astrophysical Sciences, Princeton
University, Princeton, NJ 08544, USA} 
\affil{$^2$Canadian Institute for Theoretical Astrophysics. Toronto, ON M5S3H4, Canada}

\title{On the Thermal Stability of Radiation Dominated Accretion Disks}

\begin{abstract}

We study the long-term thermal stability of radiation dominated disks
in which the vertical structure is determined self-consistently by the
balance of heating due to dissipation of MHD turbulence driven by the
magneto-rotational instability (MRI), and cooling due to radiation emitted
at the photosphere.  The calculations adopt the local shearing box
approximation, and utilize the recently developed radiation transfer
module in the Athena MHD code based on a variable Eddington tensor
rather than an assumed local closure.  After saturation of the MRI,
in many cases the disk maintains a steady vertical structure for many
thermal times.  However, in every case in which the box size in the horizontal 
directions is at least one pressure scale height, fluctuations
associated with MRI turbulence and dynamo action in the disk eventually
trigger a thermal runaway which causes the disk to either expand or
contract until the calculation must be terminated.  
 During runaway, the dependence of the heating
and cooling rates on total pressure satisfy the simplest criterion for
classical thermal instability.  We identify several physical reasons
why the thermal runaway observed in our simulations differ from the
standard $\alpha$ disk model, for example the advection of radiation
contributes a non-negligible fraction to the vertical energy flux at
the largest radiation pressure, most of the dissipation does not happen 
in the disk mid-plane, and the change of dissipation scale height with 
mid-plane pressure is slower than the change of density scale height.
We discuss how and why our results
differ from those published previously.  Such thermal runaway behavior
might have important implications for interpreting temporal variability
in observed systems, but fully global simulations are required to study
the saturated state before detailed predictions can be made.

\end{abstract}


\maketitle

\section{Introduction}
\label{sec:introduction}
The inner regions of black hole (BH) accretion disks are thought to
be radiation pressure dominated whenever the accretion rate is larger
than a few percent of the Eddington limit.  Thus, understanding the
properties of radiation dominated accretion flows is essential in
order to be able to interpret and predict the spectrum from luminous
systems such as quasars. Shortly after the standard $\alpha$ disk model
was proposed \citep{ShakuraSunyaev1973}, it was realized that radiation
dominated accretion disks are thermally unstable if the stress is
proportional to the total pressure $P_t$ \citep{ShakuraSunyaev1976}.
\cite{Piran1978} proposed a general criterion to determine thermal
stability based on the dependence of the total cooling ($Q^{-}$) and
heating ($Q^{+}$) rate on $P_t$: the disk is thermally unstable if:
\begin{eqnarray}
\left.\frac{\partial \ln Q^{+}}{\partial \ln P_t}\right|_{\Sigma} > \left.\frac{\partial \ln Q^{-}}{\partial \ln P_t} \right|_{\Sigma}, 
\label{HeatCoolrate}
\end{eqnarray}
where $\Sigma$ is the surface density. 
In a radiation dominated $\alpha$ disk model with electron scattering as the 
dominant source of opacity and stress $\tau_{r\phi} = \alpha P_t$, then 
$Q^+\propto P_t^2$ while $Q^{-}\propto P_t$ at fixed surface 
density \citep{Piran1978}, thus the inequality~(\ref{HeatCoolrate})
is satisfied and the disk is thermally unstable. 
The stability of radiation-pressure dominated disks is quite different from 
thermal instability in the much cooler disks associated with dwarf novae and 
low-mass X-ray binaries \citep[][]{Lasota2001}. The former instability is driven by strong variations in heating, 
while the latter by strong variations in the radiative cooling due to the very sharp opacity shifts near 5000K. 

There are, however, several uncertainties in the $\alpha$ disk model
which cast into doubt whether the thermal instability actually
exists in real disks.  First and foremost is the assumption that
the stress is proportional to total pressure.  For example, if
instead the stress is assumed to be proportional to gas pressure
alone (the so-called $\beta$ model), the disk is thermally stable
\citep{SakimotoCoroniti1981,StellaRosner1984,Merloni2003}.   More
importantly, it is now realized that the magneto-rotational instability
(MRI, \citealt{BalbusHawley1991}) is the physical mechanism which generates
the stress in the inner (fully ionized) regions of black hole
accretion disks \citep{BalbusHawley1991}.
Dissipation of the turbulence produced by the MRI results in heating.
However, there is no reason to expect the heating rate produced
by the MRI should be proportional to the total pressure, nor that the vertical
profiles of dissipation and heating should be the same as in an $\alpha$
disk model \citep[e.g.,][]{ZhuNarayan2013}. 
Thus, the thermal stability of MRI unstable disks is uncertain,
and can only be
investigated with nonlinear radiation MHD simulations which capture the
MHD turbulence, heating and cooling of the disk self-consistently. 
Evidence of dwarf novae thermal instability and associated limit cycle behavior 
have been found from MRI simulations \citep[][]{LatterPapaloizou2012}.

Recently, the thermal properties of radiation dominated MRI unstable disks have been
investigated in a series of papers \citep[e.g.,][]{Hiroseetal2009}, using a
module for radiation transport based on flux-limited diffusion (FLD)
developed for the ZEUS MHD code \citep[][]{Stoneetal1992I}
by \cite{TurnerStone2001}.  Using 3D local
shearing-box simulations of stratified MRI unstable disks that extend for
as long as $\sim 60$ thermal times, they report the important result that
radiation pressure dominated disks are thermally
stable \citep[][]{Hiroseetal2009}.
Moreover, they develop a simple model to
interpret this result, based on the observed time lag between stress and
pressure fluctuations which indicates that stress is not determined by
total pressure.

In this paper, we report results from a new investigation of the MHD
of radiation dominated disks using more advanced numerical methods
\citep[][]{Jiangetal2012}.  While we have been able to reproduce
many of the pioneering results reported by \cite{Hiroseetal2009}, in one
important respect our results differ: in our calculations the disk always
eventually undergoes thermal runaway (either expansion or collapse).
 Working directly with
Hirose et al., we are only able to reproduce their results if we also adopt
FLD and use the same horizontal box size as well as the initial 
condition as used in \cite{Hiroseetal2009}.  Whenever the horizontal box 
size is at least one scale height wide, or the more accurate radiation transfer 
algorithm is used, the thermal runaway shows up.
The similarities and differences between the results from the two codes will 
be reported in a future joint publication
(Jiang et al., in preparation).  In this paper, we simply report the
thermal evolution of radiation pressure dominated disks observed in our
new simulations.

This paper is organized as follows.  The equations we solve are given in
\S~\ref{sec:equations}, and the initial and boundary conditions we use
in \S~\ref{sec:INIBDcondition}.  Our primary results are
described in \S~\ref{sec:result}, while \S~\ref{sec:discussion}
presents a discussion and conclusion.

\begin{table*}[htp]
\centering
\caption{Summary of the Simulation Parameters}
\begin{tabular}{cccccccccc}
\hline
Label 		& $\Sigma/ 10^5$ g cm$^{-2}$ 	& $\rho_c$/ g cm$^{-3}$ 	& $T_c$/$10^7$ K  	&  $H$/$10^6$ cm 	& Box$/H$  			& Grids/$H$ 		& Closure & $\<P_r\>/\<P_g\>$ & Outcome\\
\hline
RHVET		& $2.15$					& 0.110				&$2.91$			&$1.46$			& $1\times20\times4$	&$32^3$			&	VET	&	8.59		&	Expand\\
RHEdd		& $2.15$					& 0.110				&$2.91$			&$1.46$			& $1\times20\times4$	&$32^3$			&	Edd	&	8.59		&	Expand\\
RMLVET		& $1.07$					& 0.0566				&$2.45$			&$1.46$			& $1\times16\times4$	&$64^2\times32$	&	VET  &	11.90	&	Expand\\
RMLEdd		& $1.07$					& 0.0566				&$2.45$			&$1.46$			& $1\times16\times4$	&$64^2\times32$	&	Edd  &	11.90	&	Collapse\\
RMEdd		& $1.07$					& 0.0566				&$2.45$			&$1.46$			& $0.45\times8.4\times1.8$	&$64^2\times32$	&	Edd  &	11.90	&	Collapse\\
RMFLD		& $1.07$					& 0.0566				&$2.45$			&$1.46$			& $0.5\times8\times2$	&$64^2\times32$	&	FLD  &	11.90	&	Fluctuate\\
RMFLDL		& $1.07$					& 0.0566				&$2.45$			&$1.46$			& $1.0\times8\times4$	&$64^2\times32$	&	FLD  &	11.90	&	Collapse\\
RSVET		& $0.253$					& 0.00212				&$2.87$			&$11.7$			& $0.4\times6.4\times1.6$	&$80^2\times40$	&	VET  &	206.45	&	Collapse\\
RSFLD		& $0.253$					& 0.00212				&$2.87$			&$11.7$			& $0.4\times6.4\times1.6$	&$80^2\times40$	&	FLD  &	206.45	&	Collapse\\
\hline
\end{tabular}
\label{Table:parameters}
\begin{tablenotes}
\item    Note: $\<P_r\>/\<P_g\>$ is the ratio of the volume and time averaged radiation and gas pressure between $20-30$ orbits after MRI saturates for RHVET and 
RMLVET. For RSVET, this number is the initial value. The Box/$H$ and Grids/$H$ are the box size and resolution for $x,z,y$ directions respectively. 
\end{tablenotes}
\end{table*}


\section{Equations}
\label{sec:equations}
We solve the equations of radiation MHD in a frame rotating 
with orbital frequency $\Omega$ at a fiducial radius $r_0$ 
from the central BH, using the local shearing box approximation 
\citep{HGB1995,Hiroseetal2009}.  Curvature of the orbit is neglected so the equations are 
solved in the local Cartesian coordinate $(x,y,z)$ with unit vectors ($\bm{\hat{i}}$, $\bm{\hat{j}}$, $\bm{\hat{k}}$),
which represent the radial, azimuthal and vertical directions respectively. 
With the vertical component of the gravitational force of the central black hole and the energy exchange 
due to Compton scattering \citep[][]{Hiroseetal2009} added, 
the equations are \citep[e.g.,][]{Jiangetal2012}
\begin{eqnarray}
\frac{\partial\rho}{\partial t}+\bfnabla\cdot(\rho \bv)&=&0, \nonumber \\
\frac{\partial( \rho\bv)}{\partial t}+\bfnabla\cdot({\rho \bv\bv-\bb\bb+{{\sf P}^{\ast}}}) &=&-\bm{ S_r}(\bP) \ \nonumber \\
+2\rho\Omega^2qx\bm{ \hat i} -\rho\Omega^2z\bm{ \hat k}-2\Omega \bm{ \hat k}\times \rho\bv,\  \nonumber \\
\frac{\partial{E}}{\partial t}+\bfnabla\cdot\left[(E+P^{\ast})\bv-\bb(\bb\cdot\bv)\right]&=&-cS_r(E)\ \nonumber \\
+\Omega^2\rho\bv\cdot(2qx\bm{ \hat i})-\Omega^2\rho\bv\cdot(z\bm{ \hat k})&-&cE_r\sigma_{sF} \frac{4(T-T_r)}{T_e},  \nonumber \\
\frac{\partial\bb}{\partial t}-\bfnabla\times(\bv\times\bb)&=&0, \nonumber \\
\frac{\partial E_r}{\partial t}+\bfnabla\cdot \bF_r=cS_r(E)&+&cE_r\sigma_{sF} \frac{4(T-T_r)}{T_e}, \nonumber \\
\frac{1}{c^2}\frac{\partial \bF_r}{\partial t}+\bfnabla\cdot{\sf P}_r&=&{\bf S_r}(\bP),
\label{equations}
\end{eqnarray}
where the radiation source terms are,
\begin{eqnarray}
{\bf S_r}(\bP)&=&-\left(\sigma_{aF}+\sigma_{sF}\right)\left[\bF_r-\left(\bv E_r+\bv\cdot{\sf P} _r\right)\right]/c \nonumber \\
&+&\bv(\sigma_{aP}a_rT^4-\sigma_{aE}E_r)/c,\nonumber\\
S_r(E)&=&(\sigma_{aP}a_rT^4-\sigma_{aE}E_r) \nonumber \\
&+&(\sigma_{aF}-\sigma_{sF})\frac{\bv}{c^2}\cdot\left[\bF _r-\left(\bv E_r+\bv\cdot{\sf  P} _r\right)\right].
\label{sources}
\end{eqnarray}

In the above equations, the shear parameter $q\equiv
-d\ln\Omega/d\ln r$ is $3/2$ for a Keplerian disk. The pressure
${\sf P}^{\ast}\equiv(P+B^2/2)\bI$ (with $\bI$ the unit tensor), and
the magnetic permeability $\mu=1$.  The total gas energy density is
$E=E_g+\rho v^2/2+B^2/2$, where $E_g=P/(\gamma-1)$ is the internal
gas energy density with $\gamma=5/3$.  The radiation constant
$a_r=7.57\times10^{15}$ erg cm$^{-3}$ K$^{-4}$.
The radiation temperature $T_r$              
is defined as $T_r\equiv \left(E_r/a_r\right)^{1/4}$, while
$T_e$ is the temperature equivalent
to the electron rest mass, $T_e=5.94\times10^9\ \text{K}$.
The frequency mean absorption and scattering opacities
(attenuation coefficients with unit cm$^{-1}$) are denoted by $\sigma_{aF}$
and $\sigma_{sF}$, while $\sigma_{aP}$ and $\sigma_{aE}$ are the Planck
and energy mean absorption opacities.

Equations \ref{equations} and \ref{sources} are solved in the mixed frame, that 
means the radiation flux $\bF_r$ and energy density $E_r$ are Eulerian variables,
while the material-radiation interaction terms in equations \ref{sources} are
written in the co-moving frame \citep[e.g.,][]{Lowrieetal1999}.  The Eulerian and co-moving flux
$\bF_{r,0}$ are related through
$\bF_{r,0}=\bF _r-\left(\bv E_r+\bv\cdot{\sf  P} _r\right)$. 
The radiation pressure ${\sf P}_r$ and energy density are related through a
variable Eddington tensor (VET),
${\sf P}_r=\bfr E_r$.  We calculate $\bfr$ from a formal solution of the transfer
equation using the method of short characteristics
\citep[][]{Davisetal2012}.  For the calculations presented in this paper,
we use $10$ angles per octant to compute the VET.

The equations are solved with the new Godunov radiation MHD code as
described and tested in \cite{Jiangetal2012} and \cite{Davisetal2012},
with the improvements given by \cite{Jiangetal2013b}. An
orbital advection scheme \citep{StoneGardiner2010} is used to speed
up the calculation. The Compton heating term is separated from other
terms and added to the energy equation in the same way as done by
\cite{Hiroseetal2009}.  These ideal MHD equations do not include 
explicit viscosity or resistivity. Instead, dissipation occurs at the grid scale 
through numerical diffusion. However, because we solve the total energy 
equation with a conservative method, the kinetic and magnetic energy dissipated 
at the grid scale is transformed into thermal energy, and is not lost.

\section{Initial and Boundary Conditions}
\label{sec:INIBDcondition}

In order to facilitate comparisons between our results, we adopt the same
values for the physical parameters as \cite{Hiroseetal2009}, as given
in Table 1 of their paper.\footnote{To correct typos in this table,
the black hole
mass we use is $1.32\times10^{34}$ g and the surface density is
twice the value given.}   The local frame of
reference represented by the shearing box
is centered at $r_0=30\left(GM/c^2\right)$, with the mass of the
central black hole $M=6.62 \msun$, and corresponding orbital frequency
$\Omega=190.1$ s$^{-1}$.  The mean molecular weight
is assumed to be $0.61$.   Following \cite{Hiroseetal2009}, we include
electron scattering opacity $0.33$ cm$^2$ g$^{-1}$, Plank-mean free-free
absorption opacity $3.7\times10^{53}\left(\rho^9/E_g^7\right)^{1/2}$
cm$^2$ g$^{-1}$ and Rosseland-mean free-free absorption opacity
$1.0\times10^{52}\left(\rho^9/E_g^7\right)^{1/2}$ cm$^2$ g$^{-1}$.

The initial vertical profile of the disk is calculated in the
same way as described in Section 2.4 of \cite{Hiroseetal2009}.
For the magnetic field, we initialize two oppositely twisted flux
tubes with the same net azimuthal flux within $-0.25< z <0.25$.
The $B_x$ and $B_z$ components of the field are generated by the
vector potential $A_B(x,y,z)=-\text{sign}(z)B_0\left[1+\cos\left(\pi
r\right) \right]/\left(32\pi\right)$ for $r\leq 0.25$, where $r\equiv
\left[x^2+\left(|z|-0.25\right)^2\right]^{0.5}$, while the $B_y$ component
is initialized from $B_y=\left(B_0^2/2-B_x^2-B_z^2\right)^{1/2}$  for
$|z|<0.8$.   This field configuration is slightly different from that
used by \cite{Hiroseetal2009} (it is symmetric with respect to the disk
mid-plane), however we have confirmed that different choices for the
initial magnetic field do not change our results on the thermal stability 
of the disk. But different initial vertical profiles can affect how fast the 
thermal runaway happens. We apply a density
floor $5\times10^{-6}$ of the initial mid-plane density throughout the
numerical integration to avoid small time steps when the density in some
cells become too low.

Shearing periodic boundary conditions \citep[][]{Jiangetal2013b}
are used for the $x$-direction, and simple periodic for the $y$-direction.
For the vertical direction, the gas pressure and density are copied to
the ghost zones from the last active cells.   The three components of
the velocity are also copied when the vertical component is outward.
To prevent inflow, they are set to zero when the vertical component is
inward.  For the magnetic field, we copy all three components to the ghost
cells when the velocity is outward, and set the horizontal components to
zero (copying only the vertical component) when the velocity is inward.
We have also tried to always copy the three components 
of magnetic field to the ghost cells even velocity is inward, 
with no effect on the results presented here.  Anomalous resistivity \citep[e.g.,][]{Sanoetal2004} 
is used within one scale height from the vertical boundary to overcome 
the numerical difficulties in the strongly magnetized photosphere 
as described in \cite{Hiroseetal2009}.
For the radiation field,
we assume zero incoming specific intensity in the short characteristics
solver used to compute the VET.
To be consistent, we set $|F_{r,z}|/\left(cE_r\right)=\sqrt{f_{zz}}$ in
the ghost zones, where $f_{zz}$ is the $zz$ component of the Eddington
tensor in the ghost zones, that is we assumes strictly outgoing
radiation flux.  The horizontal components of the radiation flux are
copied to the ghost zones from the last active zones.  The radiation
energy density $E_r$ in the ghost zones is then calculated from $\partial
(f_{zz}E_r)/\partial z=-(\sigma_{sF}+\sigma_{aF})F_{r,z}$.

\section{Results}
\label{sec:result}

From the laminar initial conditions, vigorous MHD turbulence is produced
by the MRI within a few orbits.  Thus, the central density and temperature,
as well as the vertical profiles of these quantities,
quickly evolve away from their initial values.  After saturation of the MRI,
the disk is heated by the turbulence and cooled by radiation from the photosphere.
As reported by \cite{Hiroseetal2009}, we find that after
saturation of the MRI the disk
can maintain an approximate equilibrium for many thermal times.  However,
we also find that for all the numerically converged simulations, inevitably the
disk undergoes a thermal runaway, either heating up and expanding until 
the photosphere hits the boundary of our domain, or
cooling down and collapsing until it can no longer be resolved by our
numerical mesh.  We investigate this behavior with a series
of calculations using different surface densities described below. 
Parameters and outcome of the simulations are summarized in Table 
\ref{Table:parameters}.

\subsection{Case A: Large Surface Density}

We first study a case with large surface density (twice the value
used by \citealt{Hiroseetal2009}), giving an electron scattering optical depth
across the whole disk $\tau_{\text{es}}=7.11\times10^4$.
The parameters for this run, hereafter referred to as RHVET, are  
listed in Table \ref{Table:parameters}. 
The fiducial units used in this calculation are 
$\rho_0=0.0566$ g cm$^{-3}$, $T_0=2.45\times 10^7$ K, and 
$P_0=1.89\times 10^{14}$ dyn cm$^{-2}$.
The ratio of radiation-to-gas pressure at the mid-plane initially is $4.13$.

We first ran the calculation adopting the Eddington approximation,
$\bfr=1/3\bI$, to save computer time.  The MRI saturates within the first
10 orbits, and the disk maintained a steady structure until roughly $50$ 
orbits. The total (Maxwell plus Reynolds) stress in the saturated state
normalized by the total pressure is $0.023$.
Thereafter, we observed the disk to undergo a runaway, heating up and
expanding until the total pressure increased by a factor of $\sim 6$. 
Within the first $150$ orbits of evolution, only $1.5\%$ of the total
mass is lost through the open vertical boundaries. 

\begin{figure*}[htp]
\centering
\includegraphics[width=0.9\hsize]{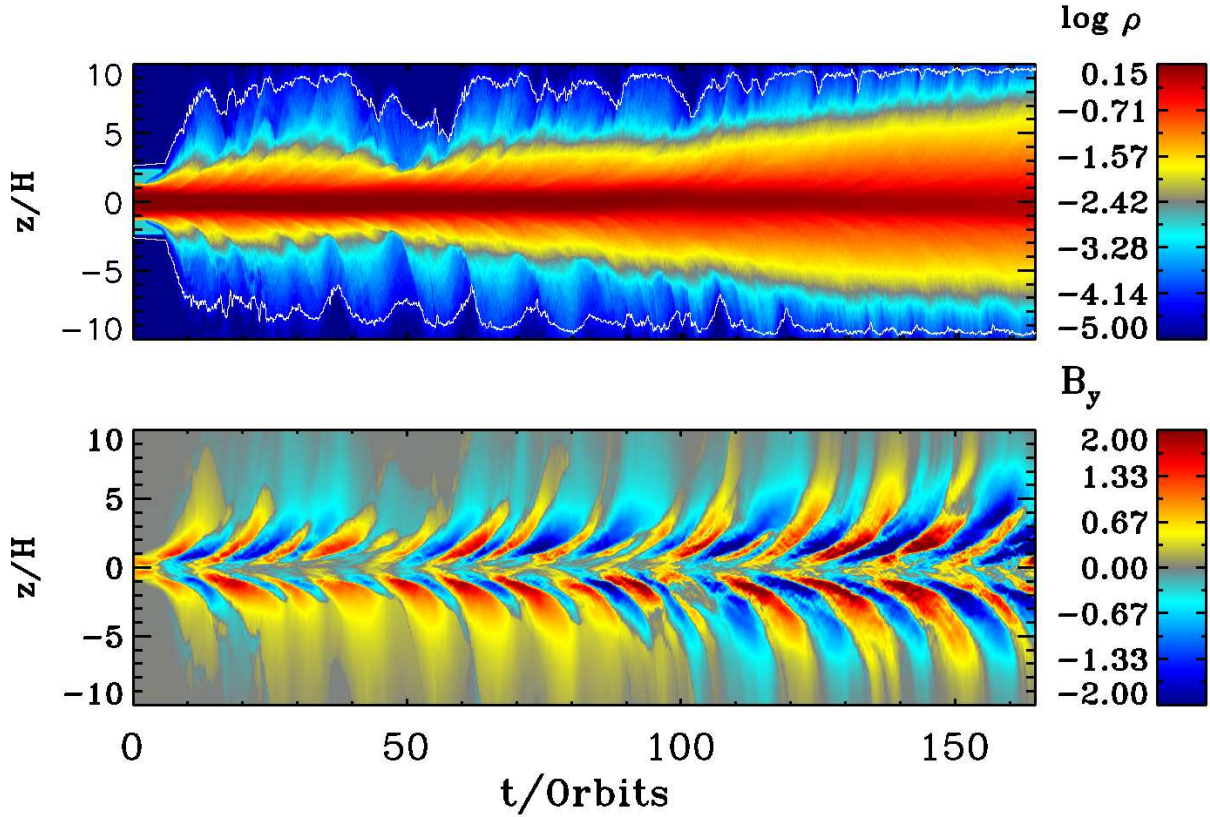}
\caption{Space-time diagram of the density $\rho$ (in unit of $\rho_0$)
and azimuthal magnetic field $B_y$ (in unit of $\sqrt{2P_0}$) for RHVET. }
\label{LarP10HST}
\end{figure*}

Concerned that this behavior is due to the approximate treatment of
radiative transfer, we restart the simulation at $50$ orbits using
a VET computed with short characteristics.  Figure  \ref{LarP10HST}
shows space-time plots of the vertical profiles of the horizontally
averaged density and azimuthal magnetic field from this calculation.
The position of the photosphere for electron scattering opacity is shown
as the white line in the top panel. It corresponds roughly to the point
at which the density drops to $10^{-4}\rho_0$.  Note the photosphere is
well inside the simulation domain for $t<100$ orbits, but has reached
the top and bottom of the box when we terminate the simulation.
The thermal time, defined as the ratio between the total energy within
the simulation domain and the 
total cooling rate, gradually increases from $10$ to $20$ orbits. 
Figure \ref{LarP10HST} shows that after an initial period of $50$ orbits, 
once again the disk undergoes a runaway over the
subsequent
$\sim 10$ thermal times. 

The space-time plot of the toroidal magnetic field 
$B_y$ shows reversals roughly every $10$ orbits, reproducing the well-known 
butterfly diagram for the MRI in stratified disks, which is observed both with
\citep{Turner2004,Hiroseetal2009,Blaesetal2011} 
and without \citep{Stoneetal1996,Davisetal2010} radiation.
This pattern is
driven by a dynamo process in the disk \citep{Brandenburgetal1995,Gressel2010,Blackman2012}. 
The photosphere of the disk also moves up and down in concert with this pattern.
Note the toroidal magnetic field fluctuations become noticeably stronger as the
disk gets thicker towards the end of the simulation.
The volume averaged Maxwell and Reynolds stress also increase secularly
as the disk is heated up, increasing by a factor of $\sim 3$ by the end of
the simulation.   This increase is consistent with the scaling of the
stress due to the MRI with domain size first noticed by \cite{HGB1995};
the larger the disk thickness, the larger the size of turbulent eddies
driven by the MRI, and the larger the stress.

\begin{figure*}[htp]
\centering
\includegraphics[width=0.9\hsize]{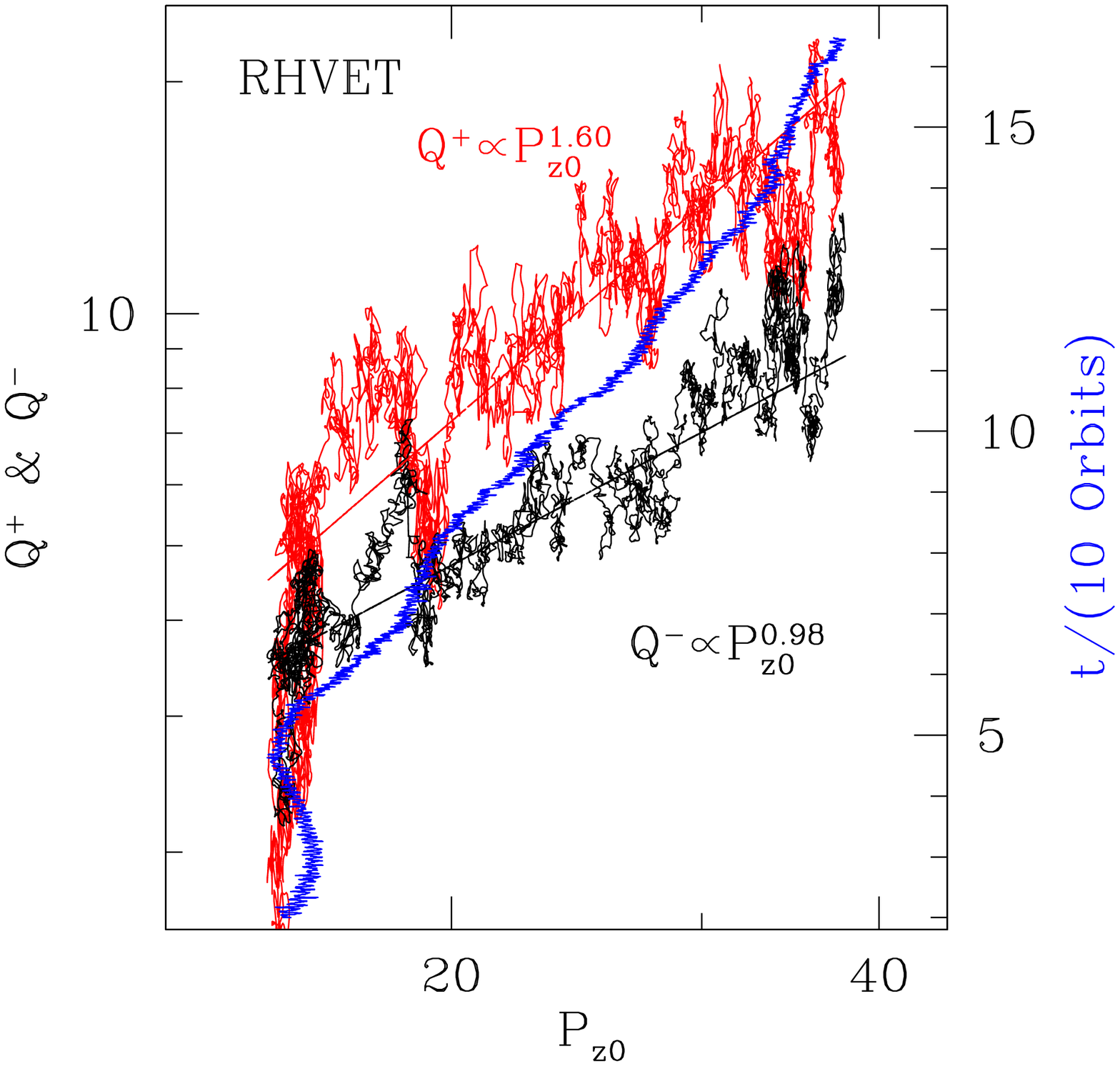}
\caption{Heating (red line) and cooling (black line) 
rates versus mi-dplane pressure for RHVET.  Also plotted is the evolution
of the mid-plane pressure with time (blue line). 
The red and black dashed lines are the best fit power-law between 
$80-130$ orbits. The unit for $P_t$ is $P_0$, while the units for $Q^+$ and $Q^-$ 
are $0.21H^3P_0\Omega$.
 }
\label{LarPVETCoolpre}
\end{figure*}

To test whether the behavior we observe is consistent with the basic tenets
of thermal instability,
we calculate the heating and cooling 
rates as follows.  The heating rate per unit area $Q^+$ in a shearing box
simulation of the MRI is set by
the work done on the fluid by the radial boundaries \citep{HGB1995,GardinerStone2005},
therefore
\begin{eqnarray}
Q^+=\frac{q\Omega}{L_y}\int_{S_x}\left(\rho v_xv_y-B_xB_y\right)dydz,
\label{heatingrate}
\end{eqnarray}
where the integration is one radial side of the domain, and 
$dy$ and $dz$ are the cell size along the $y$ and $z$ directions respectively. 
The cooling rate per unit area $Q^-$ is dominated by the radiation flux leaving
the top and bottom of the simulation domain, therefore 
\begin{eqnarray}
Q^-=\frac{1}{L_yL_x}\int_{S_z}\bF_r\cdot d\bm{A}_z. 
\label{coolingrate}
\end{eqnarray}
We plot $Q^+$ and $Q^-$ versus the mid-plane pressure $P_{z0}$,
as well as the change of 
$P_{z0}$ with time, in Figure \ref{LarPVETCoolpre}. Consistent 
with the fact that the mid-plane pressure always increases with time,
$Q^+$ is always larger than $Q^-$.  Despite large amplitude fluctuations
in $Q^+$ and $Q^-$ due to the chaotic nature of MRI turbulence, the mean values 
show a clear trend, increasing with $P_{z0}$.   We fit a power-law to both
over the period $80-130$ orbits.  During this time, 
$Q^+$ and $Q^-$ start from almost the same value, the photosphere is still well 
inside the simulation domain, and the space time plots clearly show the
thermal runaway occurring.  The best fit gives
$Q^+\propto P_{z0}^{1.60}$ while $Q^-\propto P_{z0}^{0.98}$ during this time. 
Thus, our result is consistent with the most basic premise of thermal
instability (equation \ref{HeatCoolrate}), namely the change of the 
heating rate with pressure is faster than the change of cooling rate.
However, the difference between the power law indices of 
$Q^+$ and $Q^-$ is smaller than the predicted value of 
traditional linear instability  as discussed in the introduction.

\subsection{Case B: Moderate Surface Density}

Next, we describe the results of another simulation which uses
a surface density identical to that used by  \cite{Hiroseetal2009}, that
is one half of the value used in RHVET described above, giving
an electron scattering optical depth
across the whole disk $\tau_{\text{es}}=3.56\times10^4$.  Hereafter we
refer to this simulation as RMLVET.  Parameters for this
calculation are given in Table \ref{Table:parameters}.
The fiducial units used in the calculation are
$\rho_0=0.0566$ g cm$^{-3}$, $T_0=2.45\times 10^7$ K, and
$P_0=1.89\times 10^{14}$ dyn cm$^{-2}$.
The ratio of radiation-to-gas pressure at the mid-plane initially is $4.80$.
Note that while the parameters and initial conditions for this calculation
are identical to \cite{Hiroseetal2009}, the computation is performed in
a domain which is more than two times larger in each dimension.

Once again, we first evolve the disk adopting the Eddington approximation
$\bfr=1/3\bI$ for the first $20$ orbits and restart the simulation with VET. 
After saturation of the MRI, and for the first 60 orbits of this
simulation, the total energy density and Maxwell stress fluctuate around mean
values with no systematic
trend, similar to the results reported by \cite{Hiroseetal2009}. The ratio between 
the total stress and pressure during this time is 0.034. 
However, once again after 60 orbits
the energy density and stress being to increase systematically,
and the disk expands,
albeit with large fluctuations in both during this period.
During the expansion phase, our best fit power laws to the heating and cooling
rates give
$Q^+\propto P_{z0}^{2.42}$ while $Q^-\propto P_{z0}^{1.44}$, which is again consistent with the
thermal instability criterion (equation \ref{HeatCoolrate}).

Interestingly, if we continue the simulation with the Eddington approximation,
we find the disk collapses after 80 orbits instead of expanding.


\subsection{Case C: Low Surface Density}

\begin{figure*}[htp]
\centering
\includegraphics[width=0.9\hsize]{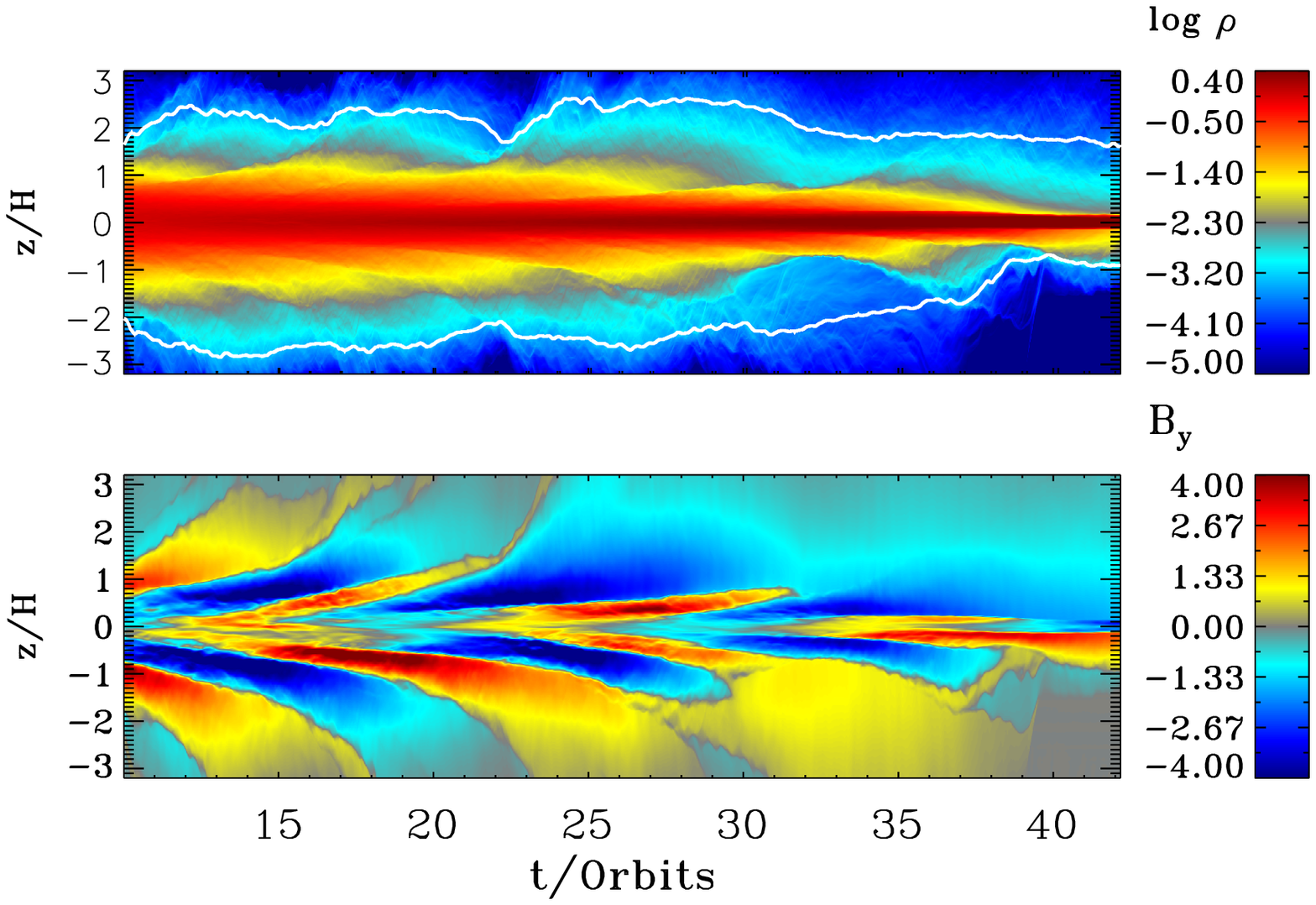}
\caption{The same as Figure \ref{LarP10HST} but for simulation RSVET. }
\label{RSVETST}
\end{figure*}

Finally, we report the evolution of a disk with low surface density,
one quarter the value used by \cite{Hiroseetal2009} giving a total
optical depth $\tau_{\text{es}}=8.36\times 10^3$.   Parameters for
this calculation, hereafter referred to as RSVET, are given in Table
\ref{Table:parameters}.  Initial parameters of this simulation are chosen
according to the radiation pressure dominated solution of the $\alpha$
disk model with assumed $\alpha=0.0125$ and $42.5\%$ of the Eddington
accretion rate.
using fiducial units
$\rho_0=0.00212$ g cm$^{-3}$, $T_0=2.87\times 10^7$ K, and
$P_0=8.29\times 10^{12}$ dyn cm$^{-2}$.
The ratio of radiation-to-gas pressure at the mid-plane initially is $206$.

We use our VET module to compute the entire evolution.  
The space-time diagram for this run after the initial 10 orbits is shown 
in Figure \ref{RSVETST}. 
For this
simulation, the stress increases for the first $20$ orbits, during
the saturation phase of the MRI, reaching a value of $0.017$ of the total pressure.
Thereafter, the stress, radiation,
and gas  internal energy densities all start to decrease.  Unlike the
previous cases, this strongly radiation pressure dominated disk does
not show any period of thermally stable structure.  After another 20
orbits of evolution, the disk has collapsed to a degree that we can no
longer resolve the MRI.  During the collapse, the best fit power laws
to the heating and cooling rates give $Q^+\propto P_{z0}^{1.90}$ while
$Q^{-}\propto P_{z0}^{0.90}$.  Therefore this solution is also consistent
with the thermal instability criterion in that when the pressure drops,
the heating rate decreases faster than the cooling rate.

\subsection{The Effects of the Radiation Transport Algorithm and Domain Size} 
\label{algorithm}

In order to assess the impact of the
radiation transfer algorithm on our results, we have implemented a FLD
module in Athena.\footnote{We note that the Athena and ZEUS
  implementations of FLD are not identical, primarily owing to the
  differences in the underlying methods (finite volume versus 
  operator splitting) and the choice of frame used to specify the radiation
  transfer equations and variables (co-moving versus mixed-frame).}  
We first consider the case of low surface density and high radiation
to gas pressure ratio.  
We restarted from a ZEUS simulation with the same parameters 
as the RSVET simulation reported in Table
\ref{Table:parameters} (kindly provided by S. Hirose),  which is labeled RSFLD. 
The initial cooling rate with FLD in Athena
is the same as in the ZEUS solution.  The disk maintains a steady
structure for $\sim 20$ orbits.  However, after $20$ orbits the
solution still collapses in a similar manner to original RSVET
simulation reported above.  If we restart the same ZEUS solution with
our VET module or Eddington approximation, the initial cooling rate is
increased by a factor of $\sim2$, indicating that FLD underestimates
the cooling rate relative to VET.  As a result, the disk cannot
maintain a steady structure and collapse proceeds immediately.
These results suggest that the Athena simulations always undergo a
thermal runaway, regardless of the radiation transfer algorithm, for
these low surface density runs where the ratio of radiation-to-gas
pressure is larger than $\sim 200$.

In domains with larger surface density, the radiation-to-gas pressure
ratio is smaller ($\sim 10$ for RMLVET).  In order to assess the
impact of the radiation transfer algorithm and the impact of domain
size in this regime, we run two simulations labeled RMEdd and RMFLD,
utilizing the Eddington approximation and the FLD algorithm,
respectively.  These use the same simulation parameters as RMLVET
except that both $L_x$ and $L_y$ are smaller by a factor of $2$, to
match the 1112a simulation of \cite{Hiroseetal2009}. We find that the
disk collapses after $\sim 150$ orbits in RMEdd.  In contrast, RMFLD
shows no clear thermal runaway over $350$ orbits, consistent with the
behavior described in \cite{Hiroseetal2009}. While the disk shows periods 
of expansion, it eventually falls back to its original structure. 
The heating and cooling rate track each other very well
during the simulation. Therefore, Athena and ZEUS agree reasonably
well for these higher surface density runs, when they utilize the same
horizontal domain size {\it and} radiation transfer algorithm (FLD).
However, when the more accurate radiation transfer algorithm is used,
the disk is still undergoes a thermal runaway even with a smaller
horizontal box size.

To further examine the effect of the horizontal box size for the case
when radiation pressure is only $10$ times the gas pressure, we
consider another run labeled RMFLDL, where we double the horizontal
box size of RMFLD run so that $L_x=H$ and $L_y=4H$ while keeping all
other parameters to be the same.  This run has the same simulation
parameters as RMLEdd and RMLVET but adopts FLD. The disk collapses
within $50$ orbits in this case, which is consistent with the thermal
runaways observed in RMLEdd and RMLVET, but different from RMFLD.
This experiment shows that by just increasing the horizontal box size
from $\sim 0.5H\times 2H$ to $H\times4H$, the disk can change from the
stable behavior to be unstable, even when FLD is used. 


The effects of the horizontal box size on the saturation state of MRI
have been studied in previous isothermal MRI
simulations. \cite{Simonetal2012} studied the saturation of MRI
through a series of simulations with different horizontal box
sizes. They found that the fluctuation of MRI turbulence decreased
with increasing horizontal box size.  Furthermore, properties of the
MRI turbulence, such as vertical dissipation profile and correlation
length, are only converged when the horizontal box size is larger than
$H\times 4H$.  It has also been observed that small box MRI
simulations shows large spikes in stress, and thus heating rate, due
to the recurrent channel solutions, which are significantly reduced
when the horizontal box size is increased \citep[][]{Bodoetal2008,
  Jiangetal2013b}.   Larger box size also increases the number 
  of modes in the system and phase trajectories of the system are 
  less constrained.  These results suggest that the simulations with
larger domains are more reliable to determine whether the radiation 
dominated disks are thermally stable or not. 


These results indicate the outcome of the simulations depends on the
radiation-to-gas pressure ratio, the box size, and the radiation
transport algorithm used.  For modest radiation-to-gas pressure ratios
($\sim 10$), the box size and the algorithm used both have an impact,
although our preferred setup (larger domains with the VET algorithm)
produces thermal runaway.  When the radiation pressure is $\sim 200$
times the gas pressure and the surface density is small, the
differences caused by the box size and radiation transfer algorithm
are less important and we observe a thermal runaway for all the
simulations.


\section{Discussion}
\label{sec:discussion}

The final outcome of all the simulations reported above is summarized
in Table \ref{Table:parameters}.  We will report a detailed analysis of
these simulations, as well as the results from many more that explore
a much wider range of radiation-to-gas pressures, in a future paper
(Jiang et al., in prep.).  Here, we first show that although we 
find thermal runaway as classical thermal instability predicted, 
many assumptions in the $\alpha$ disk model are not satisfied in our 
simulations and the details of the thermal runaway differ from
the predictions.Then we discuss how our radiation
dominated results compare to simulations run with the ZEUS code.




\subsection{Comparison with the $\alpha$-disk Model}

Before we begin an in-depth comparison, 
it is useful to state precisely what is meant by the $\alpha$-disk model in
the current context.  In terms of vertical structure, the
$\alpha$-disk is effectively a ``one-zone'' model as all variables are
characterized by a single value at a given radius.  Since we are
comparing with local simulations with fixed mass, it is useful to
consider the properties of the disk at a single radius, with
constant values of surface density $\Sigma$, $\Omega$, and
surface radiation flux $F_{\rm rs}$.  One to one correspondence is
assumed between stress and the total mid-plane pressure $\tau_{r\phi}
= \alpha P_{z0}$, with $\alpha$ constant.  We focus here on the
radiation dominated limit where $P_{r}(0) \equiv P_{z0}$.

The disk is assumed to be in hydrostatic and thermal equilibrium,
which given $\Sigma$, $\Omega$, and $F_{\rm rs}$ determines the
mid-plane radiation pressure via
\begin{eqnarray} 
F_{\rm rs} =\frac{2 \xi c P_{z0}}{\kappa_{\rm s} \Sigma}.
\end{eqnarray}
Here $\kappa_{\rm s}$ is the electron scattering opacity and 
$\xi$ is a parameter that depends on the vertical distribution
of dissipation, which is assumed to be constant in the $\alpha$-disk
model.  If one follows \citep{ShakuraSunyaev1973} and assumes $d F_r/
d z \propto \rho$, $\xi = 1/2$.  The cooling rate per unit area
in the disk is assumed to be determined solely by radiative cooling
so that $Q^{-}=F_{\rm rs}$.
The heating rate per unit area is proportional to an integral over $z$ 
of the stress
\begin{eqnarray}
Q^{+} = \frac{3\Omega}{2} \int \tau_{r\phi} dz \simeq
 \frac{3}{2} H_{\rm s} \tau_{r\phi} \Omega.
\end{eqnarray}
with $H_{\rm s}$ defining a characteristic scale height for the stress.
Finally, hydrostatic equilibrium determines a characteristic flux
scale height
\begin{eqnarray}
H_{\rm F} = \frac{\kappa F_{\rm rs}}{c \Omega^2} \simeq \frac{2 \xi P_{z0}}{\Omega^2 \Sigma}.
\end{eqnarray}
Thermal instability follows from equation \ref{HeatCoolrate} by
assuming $\xi$ and $\alpha$ are constants and that $H_{\rm s}=H_{\rm
  F}$.  We then find that $Q^{+} \propto P_{z0}^2$ and $Q^{-}
\propto P_{z0}$.  This simple model then predicts a linear instability
that grows on the order of the thermal time $\simeq 1/(\Omega \alpha)$.

Strictly speaking, {\it none} of these assumptions are obeyed by the
shearing box simulations, therefore it is not surprising that we do not find
an exponential runaway on the thermal timescale. For the moderately
radiation pressure dominated case, the disk does in fact stay in a
roughly equilibrium state with large amplitude fluctuations for
several thermal times.  Both RHVET and RMLVET survive for $\sim 50-60$
orbits before thermal runaway takes over.  In fact, it is not even
clear how a classical linear analysis can be applied to an MRI
turbulent disks, because the disk already contains nonlinear amplitude
fluctuations in the stress, heating and cooling rates driven by the
turbulence.  Even the assumption of a single equilibrium state is
questionable because variations in the vertical distributions of
dissipation and stresses can, in principle, offer a range of
equilibrium (or near-equilibrium) configurations to the turbulent flow.

Instead, the behavior we observe is perhaps more akin to a nonlinear
instability, in which a finite amplitude fluctuation is required to drive
the system away from equilibrium.  In this interpretation, the equilibrium
state in the radiation pressure dominated case may be better described
as a basin of attraction which temporarily traps solutions.  Even when
there is no runaway, turbulence provides large amplitude fluctuations in
the disk structure which causes the system to wander within the 
basin \citep[][]{CovasAshwin1997,AshwinRucklidge1998,JaniukMisra2012}.
Exploring the topology of solutions in phase space for this dynamical
system to discover what conditions fluctuations must satisfy to produce
runaway is well beyond the goals of this paper. Nonetheless, such an
interpretation may provide an explanation for the effects of horizontal 
box size on the thermal instability we find.  

We can quantify the difference with the $\alpha$-disk by computing some
appropriately averaged quantities.  We start by computing the
scaleheight $H_{\rm s}$
\begin{eqnarray}
H_{\rm s}=\frac{\int \int \int |z| \tau_{r\phi} dx dy dz}{\int \int \int\tau_{r\phi} dx dy dz},
\end{eqnarray}
where all integrals are performed over the full extent of the domain.

\begin{figure*}[htp]
\centering
\includegraphics[width=1.0\hsize]{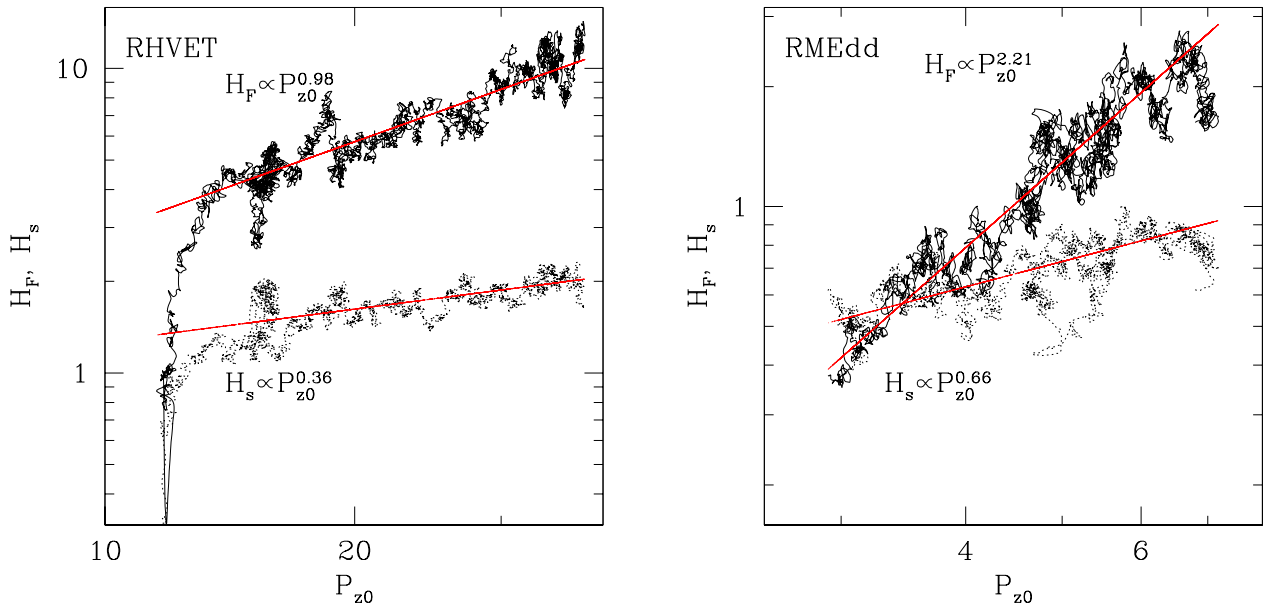}
\caption{The flux (solid) and stress (dotted) scale heights versus total mid-plane
pressure for simulation RHVET (left panel) and RMEdd (right panel). 
The best fit power law relations are shown as red lines.
}
\label{scaleheight}
\end{figure*}

We plot $H_{\rm s}$ and $H_{\rm F}$ in Figure \ref{scaleheight} for
the simulations RHVET and RMEdd versus the total mid-plane pressure.  
We see that
$H_{\rm F}$ is not directly proportional $P_{z0}$ due to
fluctuations.  It is also generally larger than the $\alpha$ model
prediction (especially in simulation RMEdd) 
because the dissipation rate per unit mass from MRI
turbulence actually increases with decreasing column density. 
This leads to more dissipation near the surface and a lower ratio of
surface flux to mid-plane pressure.  Especially when the 
density profile changes significantly with mid-plane pressure, 
$H_{\rm F}$ deviates from linear dependence on $P_{z0}$ 
significantly.
 The red lines shows best-fit
power-law relation between the scale heights and pressure, fit from
80-130 orbits for RHVET and 20-150 orbits for RMEdd. 
In the RHVET run, this is close to linear for $H_{\rm
  F}$, suggesting that the profile of the dissipation per unit mass is
relatively constant (at least when averaged over fluctuations), in
agreement with the $\alpha$-disk assumption.  However, 
simulation RMEdd shows a tendency for a steeper than linear
increase in $H_{\rm F}$ with $P_{z0}$, indicating a a larger share
of the dissipation occurs at low densities as the disk expands.  We
also see that $H_{\rm s}$ differs significantly from $H_{\rm F}$,
particularly at large pressure, due to a much weaker than linear
increase with pressure ($H_{\rm s} \propto P_{z0}^{0.36}$ for
RHVET and $H_{\rm s} \propto P_{z0}^{0.66}$ for RMEdd). 
This is a fairly general result that holds in almost all of
our simulations.

\begin{figure}[htp]
\centering
\includegraphics[width=1.0\hsize]{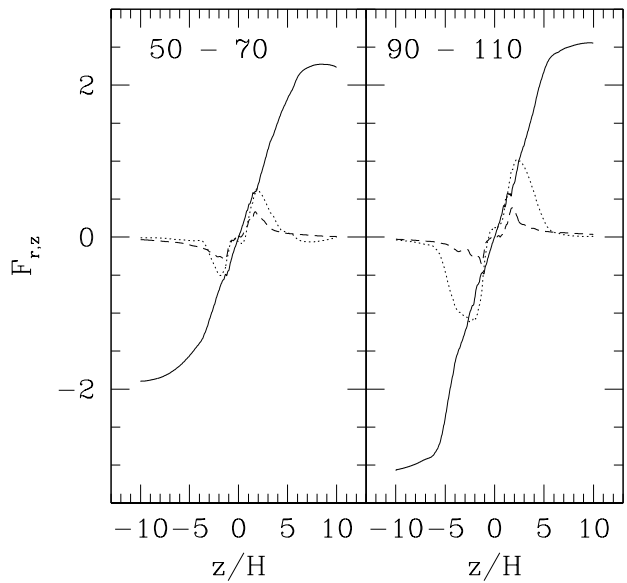}
\caption{Time and horizontally averaged vertical profiles of diffusive 
(solid line), advective (dotted line) and Poynting flux (dashed 
line) for RHVET at two different times. The left panel shows averages over
$50$ and $70$ orbits, while the right panel shows averages over
$90-110$ orbits. The unit of flux is $\sqrt{P_0/\rho_0}P_0$.
}
\label{LarPVETAdvection}
\end{figure}

The different dependence of $H_{\rm F}$ and $H_{\rm s}$ on pressure is
closely related to the presence of vertical fluxes of energy other
than radiative diffusion.  During thermal runaway, the advective flux
of radiation energy near the mid-plane due to turbulence and magnetic
buoyancy \citep{Blaesetal2011} is not negligible, as is assumed in the
$\alpha$ disk model.  Figure \ref{LarPVETAdvection} shows vertical
profiles of the temporally and horizontally averaged diffusive,
advective, and Poynting flux for RHVET at two different times.  Both
the advective and Poynting flux peak at $\sim 2H$.  Thereafter both
drop quickly to zero approaching the photosphere.  The ratio between
the advective and diffusion flux near the mid-plane becomes larger
when the disk becomes hotter.  At $90$ orbits, the advective flux is
larger than the diffusive flux within $\sim 2H$.  The diffusive flux
is constrained by the vertical component of gravitational
acceleration, which does not change, whereas the advective flux
depends on the amplitude of the turbulence and magnetic energy (which
determines buoyancy), both of which are increasing as the disk gets
thicker.  Note that at the photosphere, radiation is indeed the
dominant cooling mechanism of the disk, as assumed in the $\alpha$
model: at the top and bottom of the simulation domain, the mechanical
and Poynting fluxes are only $\sim 1\%$ of the radiation flux in
RHVET.  Because of the extra cooling provided by the advective flux at
the mid-plane, the radiative flux, and therefore, $H_{\rm F}$ can
increase without a proportional rise in $H_{\rm s}$.


The estimation of $\alpha$ is somewhat more arbitrary.  We could,
in principle, compute $\alpha$ as the ratio of the volume averaged
stress to the volume averaged total pressure, but this would be
contrary to the $\alpha$ model assumption, which implicitly scales $\alpha$
with the mid-plane pressure.  The distinction is relevant because the
vertically averaged pressure generally scales as the square of the
mid-plane pressure.  Simply using the ratio of the mid-plane stress and
pressure is also problematic because the stress and pressure show
markedly different profiles, even when time averaged.  While the
pressure peaks at or near the mid-plane, the stress peaks somewhat off
the mid-plane with a local minimum at the mid-plane.  Hence we prefer to
define $\alpha = \langle \tau_{r\phi} \rangle/P_{z0}$,
where $\langle \tau_{r\phi} \rangle$ is the volume average
of the stress for $|z| < H_{\rm s}$.

For this definition, we find that $\alpha$ is generally a weakly
increasing function of mid-plane pressure, but one with large
fluctuations about the best-fit mean relations.  For example, the
RHVET simulation provides a best-fit $\alpha \propto P_{\rm
  t0}^{0.19}$.  The this trend of weakly increasing $\alpha$ with
pressure is a fairly general result in our simulations, and is in
approximate agreement with the $\alpha$ model assumption.  Along with
the weak increase of $H_{\rm s}$ with pressure, this implies a value
of $\partial \ln Q^{+}/\partial \ln P_{z0} < 2$ for most
simulations.  In the specific case of RMFLD simulation, which shows no
evidence for runaway, the dependence is sufficiently weak that
$\partial \ln (Q^{+}/Q^{-})/\partial \ln P_{z0}$ is only about
0.1.


Despite the differences discussed above, we have shown that during
runaway our solutions agree with the most basic criterion for thermal
instability: the heating rate $Q^+$ has a steeper dependence on 
mid-plane pressure $P_{z0}$ than does the cooling rate $Q^-$, albeit with
significant fluctuations on shorter timescales.  However, the actual
scalings we measure can be substantially different from what the
$\alpha$ model predicts.  In particular, the difference in the
logarithmic slopes of $Q^+$ and $Q^-$ is generally smaller because the
scaling of $Q^+$ with $P_{z0}$ is weaker than the $\alpha$ model
predicts due primarily to a weaker than expected scaling of $H_{\rm
  s}$ with $P_{z0}$.

\subsection{Comparison with ZEUS Results}

Prior to this work, radiation MHD simulations performed in the
radiation dominated regime have only been run with the ZEUS FLD
algorithm \citep[e.g.][]{Turner2004,Hiroseetal2009}.  In all cases,
the simulations were reported to be consistent with thermal stability,
in tension with our Athena results, for which thermal runaways are the
norm.   \citet{Hiroseetal2009} explained this stability by noting that 
fluctuations in the magnetic field lead those in the radiation. And 
therefore they argue that magnetic field controls the radiation, not 
vice versa. 


The picture outlined above also applies to the Athena simulations.  In
fact, we see an almost identical time lag between the fluctuations of
radiation pressure and magnetic energy, with magnetic pressure leading
the radiation pressure by $\sim 10$ orbits.  Therefore, differences in
the stability properties of the two codes most likely originates in
differences in the way the radiative cooling and the saturation level
of the turbulent stresses respond to fluctuations in the pressure. The
latter topic is still not well understood, even in isothermal
stratified simulations, and requires a direct and detailed comparison
between the codes.  In collaboration with Hirose et al., we have
initiated a detailed comparison between the codes and the simulations
results, which will be the subject of a future paper (Jiang et al. in
prep.).  Since the focus of this paper is the Athena results, we
simply summarize the most salient area of conflict or agreement
between the simulations.

Overall, our solutions confirm many important results regarding the
vertical structures and time evolution reported with ZEUS. As
discussed in the previous sections, our simulations can run for many
thermal times without any clear evidence of thermal runaway.  In
particular, we note that during the first 50 orbits of evolution in
the moderately radiation dominated domains (RHVET and RMLVET), we find
a relatively steady vertical structure is maintained for several
($\sim 5$) thermal times, characterized by large amplitude
fluctuations in the stress and energy densities but with no secular
trends.  When a thermal runaway does eventually occur, it generally
takes many thermal times for the expansion or contraction to
proceed. Hence there is no sign of an exponentially growing
fluctuations as one would expect from a linear instability.  

The discrepancies between the results only becomes apparent on long
time scales (i.e.  many thermal times), since the
\cite{Hiroseetal2009} simulations do not show clear evidence of
runaway behavior, even when evolved for much longer times (up to $600$
orbits).  The results of section \ref{algorithm} suggest that these
discrepancies may be attributed, at least in part, to both the
horizontal domain size and the use of different radiation transfer
algorithms.  It is reassuring that the simulations seem to agree
when we use Athena's FLD algorithm with the same domain size and
resolution as ZEUS.

In contrast, Athena simulations always end in a thermal runaway at
the larger radiation-to-gas pressure ratios ($\sim 200$ in RSVET and
RSFLD), independent of the transfer algorithm employed. 
Recently, experiments using ZEUS in the large radiation pressure, low surface 
density regime as in RSVET (Hirose et al., private communication 2013), 
find similar behavior. 
 They perturbed the disk temperature by $2\%$ to $10\%$ after restarting 
evolution of a ZEUS simulation with the same parameters as RSVET at $100$ 
orbits.
Depending on the type of the
perturbations they added, the disk expanded, collapsed, or continued running in
an apparently stable equilibrium. When they continued to run the simulation to 
300 orbits without adding any perturbation, the disk still collapsed. 
Therefore, it seems possible for ZEUS to find similar
thermal runaways to those described above when the domains are
very radiation dominated.

\section{Conclusions}

The evolution of MRI-unstable radiation pressure dominated accretion
disks computed with the new radiation-transfer module in Athena
always show thermal runaway. The disk can either expand or collapse,
depending on the surface density of the disk, and numerical parameters
of the calculations.  The dependence of the heating and cooling rates,
$Q^+$ and $Q^-$ respectively, on mid-plane pressure is consistent with the
general criterion for thermal instability, in that $Q^+$ increases with
mid-plane pressure faster than $Q^-$.  

For the strongly radiation pressure dominated case with lower surface
density, the disk always undergoes a thermal runaway, independent
of the radiation transfer algorithm and the numerical parameters. The
time lag between pressure and magnetic energy density fluctuations,
which is used to explain the results in \cite{Hiroseetal2009}, still
exists in all these unstable solutions.  This suggests that the time
lag itself is not sufficient to stabilize the disk, which is
consistent with the linear analysis including the time lag
\citep[][]{Ciesielskietal2012}.

For the marginally radiation pressure
dominated case, the ZEUS code does not show thermal runaway behavior
reported here even over much longer integration times (up to $600$
orbits).  To investigate this difference, we have implemented FLD in
Athena. We are able to reproduce the behavior reported by
\cite{Hiroseetal2009}, only if we use FLD and the small horizontal box
size ($\sim L_x=0.5H, L_y=2H$) in Athena. When we use different
radiation transfer algorithms for this small box size, or we increase
the horizontal box size in Athena, we always find a thermal
runaway. Therefore, the previous reports of thermal stability in the
moderately radiation pressure dominated disk may be because of, in part,
the smaller box size used. We believe simulations run on larger domains to 
be more robust \citep[e.g.,][]{Simonetal2012}.


A natural question about the observed expansion and collapse in our
simulations is whether this behavior will continue if a larger vertical
box size, or higher numerical resolution at the mid-plane, are used.
For example, in those cases where the disk collapses, we cannot resolve
the MRI modes that may still fit into the disk, and it is possible these
modes might eventually lead to turbulence, heating, and re-expansion of the disk.
However, in many respects this issue is moot.  The fact that the mid-plane
pressure can change by a factor of $\sim 10$ has important consequences
for the structure, evolution, and observational appearance of radiation
dominated disks.  Moreover, if the vertical thickness is varying by such
large amplitudes in local patches of the disk, we expect radial fluxes
of mass, momentum, and energy may become important, which invalidates the
use of the shearing box approximation.  Thus, we consider it appropriate to
characterize fluctuations of such large amplitude as a runaway, and moreover
the final state reached in such runaways
is clearly beyond the scope of local shearing box simulations,
and requires global models.

All the simulations reported here are done with net azimuthal
magnetic field.  It would be very interesting to see how the radiation
dominated disks behave with net vertical magnetic flux. MRI with net
vertical magnetic flux can generate more vigorous turbulence than the
net azimuthal magnetic field case and the stress increases with
vertical box size in the unstratified simulations \citep{HGB1995}.
Therefore, we expect the stress will also increase with disk scale
height and thermal instability should still exist in this
case. However, future simulations are required to confirm
this supposition


%

The thermal runaways we find here may have important implications for 
observations. The $\alpha$ disk model predicts that the classical thermal 
instability will lead to a limit-cycle behavior, where the disk switches 
between a low temperature, gas pressure dominated state and high 
temperature, radiation pressure dominated state~\citep[][]{Janiuketal2002,Doneetal2007}. 
However, observations of most Galactic X-ray binaries do not find 
variability similar to that predicted \citep[][]{GierlinskiDone2004}, 
except for the well known source GRS1915+105 \citep[][]{Doneetal2004,Doneetal2007} 
and recently reported source IGR J17091-3624 \citep[][]{Altamiranoetal2011}. 
It is believed that the accretion rate
reaches the Eddington limit in GRS1915+105, and this is the
trigger for instability in this case
\cite[][]{Doneetal2004,Doneetal2004b}. 
Our new simulations show that although thermal runaway may occur, it does
so on growth rates that are smaller by a factor of a few compared
to the $\alpha$ model. 
Moreover, when the radiation 
pressure is only a few times the gas pressure (as in RHVET and RMLVET), the disk can survive
for several thermal times before runaway occurs.  In this case 
the classical limit-cycle behavior will likely be modified.  On the other hand,
when the radiation pressure is hundred times the gas pressure (as in RSVET),
the disk collapses within $\sim 2$ thermal times. In this case, we expect 
that the classical limit-cycle behavior may emerge
\citep{Janiuketal2002}, which may have relevance to the observed variability of 
GRS1915+105. 
  
Finally, we remark that
the observational 
appearance of real disks will depend on how the thermal runaway saturates \citep[][]{DexterQuataert2012}.
This cannot be addressed in the local shearing box approximation.  Moreover,
the thermal instability can be modified by radial advection of energy
\citep{Abramowiczetal1988}, and
by fluctuations in the surface density on long wavelengths 
\citep{LightmanEardley1974}.  All of these
issues require global simulations of the saturation of the MRI in radiation
dominated disks, which is the focus of our current effort.

\section*{Acknowledgements}
We thank Omer Blaes, Shigenobu Hirose, and Julian Krolik for very
helpful discussions, and giving us restart dumps from their
simulations. We also thank Jeremy Goodman for valuable comments on the
results. The anonymous referee also gives us helpful comments to
improve the paper. This work was supported by the NASA ATP program
through grant NNX11AF49G, and by computational resources provided by
the Princeton Institute for Computational Science and Engineering.
Some of the simulations are done in the Pleiades Supercomputer
provided by NASA.  This work was also supported in part by the
U.S. National Science Foundation, grant NSF-OCI-108849.  Some
computations were also performed on the GPC supercomputer at the
SciNet HPC Consortium. SciNet is funded by: the Canada Foundation for
Innovation under the auspices of Compute Canada; the Government of
Ontario; Ontario Research Fund - Research Excellence; and the
University of Toronto.  S.W.D. is grateful for financial support from
the Beatrice D. Tremaine Fellowship.

\bibliographystyle{apj}
\bibliography{ThermalStability}
\end{CJK*}

\end{document}